\documentclass{pazhb}
\usepackage[hyperfootnotes=false]{hyperref}
\usepackage{color}
\usepackage{epsf}
\usepackage[utf8]{inputenc}
\usepackage[russian]{babel}
\usepackage[T2A]{fontenc}
\usepackage{amsmath}
\usepackage{amsfonts}
\usepackage{amssymb}
\usepackage{epsf}
\usepackage{graphicx}
\usepackage{gensymb}
\usepackage{float}
\usepackage{longtable}
\usepackage{changes}
\usepackage{wrapfig}

\usepackage{indentfirst}   

\begin{document}

\newcommand{\obj}{eXV3~4638}

\journalinfo{2022}{0}{0}{1}[0]
\UDK{///}

\title{SRGe2149+6736 ---  the new candidate to AM~Her type variables discovered by eROSITA telescope on "Spectrum--Roentgen--Gamma"\, orbital observatory\, }
\author{%\bf \hspace{-1.3cm}\copyright\, 2020 г. \ \
I. F.~Bikmaev\address{1,2,7}\email{ibikmaev@yandex.ru},
A. I.~Kolbin\address{5,1},
V. V.~Shimansky\address{5},
I. M.~Khamitov\address{6,1}, 
E. N.~Irtuganov\address{1,2,7}, E. A.~Nikolaeva\address{1,2,7},
N. A.~Sakhibullin\address{1,2}, R. I.~Gumerov\address{1,2}, R. A.~Burenin\address{3}, M. R.~Gilfanov\address{3,4}, I. A.~Zaznobin\address{3},
R. A.~Krivonos\address{3},
P. S.~Medvedev\address{3}, A. V.~Mescheryakov\address{3}, S. Yu.~Sazonov\address{3}, R. A.~Sunyaev\address{3,4},  G. A.~Khorunzhev\address{3},
A. V.~Moiseev\address{5,3},
E. A.~Malygin\address{5},
E. S.~Shablovinskaya\address{5},
S. G.~Zheltoukhov\address{7}
\addresstext{1}{Kazan (Volga-region) Federal University, Kremlevskaya 18, Kazan 420008, Russia}
\addresstext{2}{Academy of Sciences of Tatarstan Rep., Baumana 20, Kazan 420111, Russia}
\addresstext{3}{Space Research Institute of RAS, Profsoyuznaya 84/32, Moscow 117997, Russia}
\addresstext{4}{Max Planck Institute for Astrophysics, Karl Schwarzschild Str. 1, Garching, Germany}
\addresstext{5}{Specal Astrophysical Observatory of RAS, Nizhnij Arkhyz,  Karachai-Cherkessian Rep., 369167, Russia}
\addresstext{6}{TÜBİTAK National Observatory, Antalya, Turkey}
\addresstext{7}{Sternberg Astronomical Institute of M.V. Lomonosov Moscow State University, Moscow 119234, Russia}
}

%\vspace{2mm}
%\received{\today}
%\sloppypar
%\vspace{2mm}
%\noindent

\shortauthor{}

\shorttitle{SRGe2149+6736}

\begin{abstract}
We present the results of the optical identification, classification, as well as analysis of photometric and spectral observations of the X-ray transient SRGe2149+6736 detected by the eROSITA telescope during SRG all-sky X-ray survey. Photometric observations of the optical companion of SRGe2149+6736 were carried out on 6m telescope BTA SAO RAS, 1.5m Russian-Turkish telescope RTT-150 and 2.5m telescope CMO of Moscow State University.  Together with ZTF data they showed that the source is a cataclysmic variable with an orbital period  $P=85\pm0.4$~min which demonstrates long-term brightness variability from $23.5$~mag (low state) to $20$~mag (high state).  The high-state light curves are consistent with a model of accreting magnetic white dwarf and suggest that SRGe2149+6736 belongs to AM~Her type variables. The optical spectra obtained in the low state are consistent with a spectral energy distribution of a white dwarf with a temperature of $\sim 24000$~K. 

\end{abstract}

%***************************************************************
\section*{INTRODUCTION}
\noindent

The Spectrum--Roentgen--Gamma (SRG) X-ray observatory \citep{sunyaev2021} launched July 13, 2019 successfully works on an orbit around the Lagrangian point L$_2$ of the Sun-Earth system.  The main purpose of the observatory is a survey the entire sky in a wide range of energies 0.3--30~keV. Four full sky surveys completed by mid-December 2021. The observatory includes two X-ray telescopes with glancing incidence mirrors: the SRG/eROSITA operating in the 0.3--10~keV range \citep{predehl2020} and the SRG/ART-XC of M.N. Pavlinsky, 5--30~keV range \citep{Pavlinsky2021}). 
As a result of four sky surveys, the SRG/eROSITA telescope has recorded more than two million X-ray sources in the half of the sky where Russian scientists are responsible for processing the data. The vast majority of the registered sources are active galactic nuclei (AGN) and quasars. In addition to the stationary sources registered in each scan, the SRG/eROSITA telescope recorded variable X-ray sources from 2019 to 2022. Their flux varied more than 7 times between scans or was recorded only in one scan. Such variable sources include the phenomena of tidal destruction of stars in the vicinity of active galaxy nuclei \citep{Sazonov2021}.

As a result of optical observations, some of such sources turned out to be cataclysmic variables with variability in the X-ray range. The SRGe2149+6736 source studied here is also one of these sources.
The SRGe2149+6736 source was detected in the 4th SRG/eROSITA scan in July 2021. The infrared source \textit{J214918.91+673633.3} is located in the 98\% localization region with radius $R98 = 5.5''$ around the X-ray coordinates of SRG/eROSITA according to archived data from the CatWISE2020 Catalog \citep{marocco21}.
At a distance of $0.3''$ from the infrared source an optical transient \textit{ZTF21ablobhh} with coordinates RA~$=327.32858^{\circ}$, DEC~$=+67.609050^{\circ}$ was registered on 12 July 2021 by the ZTF \citep{masci18}. The transient had brightness $r=20.1\pm0.2$~mag.

The phometric observations carried out on the BTA in the $i$-SDSS filter detected an optical companion in the localization region, which coincides in coordinates with ZTF21ablobhh. Fig. ~\ref{fig:findingchart} shows the position of the found source. The brightness of the source clearly varied between neighboring exposures. A long photometric series obtained on the 1.5-m telescope RTT-150 showed brightness variability typical for magnetic cataclysmic variables. To determine the parameters of this optical candidate for SRGe2149+6736 (hereafter referred to as e2149 for brevity), we carried out a series of photometric and spectral observations on the 1.5m telescope RTT-150, the 2.5m telescope of the CMO GAISh MSU and the 6-m telescope BTA of SAO RAS. The results of these observations are analyzed in this paper.

Cataclysmic variables are close binary systems consisting of an accreting white dwarf (primary) and a main-sequence star (secondary) filling its Roche lobe and losing matter through the vicinity of the Lagrangian point L$_1$ \citep{Warner95}. The orbital periods of these systems lie in the range from $\approx 82$~min \citep{Knigge11} to several hours.

Accretion in cataclysmic variables depends on the magnetic field strength of the white dwarf. In systems with weak magnetic field of white dwarf ($B\lesssim 0.1$~MG) an accretion disk is formed. Many members of such type exhibit periodic outbursts with an amplitude $\Delta V = 2-6$~mag and are called dwarf novae. The outbursts arise due to the thermal instability of the accretion disk that occurs at low accretion rates ($\dot M\lesssim 10^{-9}$~M$_{\odot}$/year). Outbursts of dwarf novae last from a few days to a few months and can recur on time scales of weeks to tens of years. Another type of cataclysmic variables are novalikes which do not exhibit outbursts. These systems are thought to have a high accretion rate ($\dot M\gtrsim 10^{-9}$~M$_{\odot}$/year) at which the disk is stable. Related to cataclysmic variables are AM~CVn type stars with degenerate donors \citep{Solheim10}. They are characterized by short orbital periods ($P_{orb}=5-65$~min) and the absence of hydrogen lines in the optical spectra.

Accretion disk isn't formed when the white dwarf is strongly magnetized ($B\sim 10-200$~MG). The ionized gas in the accretion stream quickly reaches the Alfven radius and flows along the magnetic field lines toward the magnetic poles.  Systems of this type are called AM~Her variables or polars. A strong magnetic field in polars synchronize the rotation of the white dwarf with its orbital motion ($P_{spin} = P_{orb}$). The matter falling on the accretor forms hot ($T \sim 10-50$~keV) accretion spots on the surface of the white dwarf which are sources of X-ray radiation and cyclotron radiation in the optical range. We refer the reader to the review of \cite{Cropper90} for a more detailed introduction to AM~Her type systems. Systems with weaker white dwarf magnetic fields ($B\sim 0.1-10$~MG) are referred to the DQ~Her type or intermediate polars \citep{Patterson94}. Accretion disks can form in these systems, but their inner parts are destroyed by the magnetic field of the white dwarf. Unlike polars, intermediate polars have not synchronization of the rotational and orbital motion of the white dwarf (the average ratio of rotation period to orbital one $\langle P_{spin} / P_{orb} \rangle \approx 0.1$). 

\section{SRG/eROSITA X-ray observations}

The source e2149 was detected by the SRG/eROSITA telescope in 4 sky surveys and observed in passages between 2021/07/16 and 2021/07/19.
X-ray source coordinates are RA~$=327.33046^{\circ}$ and DEC~$=67.60941^{\circ}$, X-ray source localization error is $R98=5.5''$. 
The fourth survey recorded 44 X-ray photons from the source in the 0.3--2.2~keV range.
The average flux of the source in the fourth survey is $(11.2 \pm 2) \times 10^{-14}$~erg/s/cm$^2$. 
Interestingly, the eROSITA telescope observed the source region only 4 days after the bright optical flare ZTF21ablobhh, and at this location detected a bright source now in the X-ray range.

The source was not detected in the previous sky surveys of SRG/eROSITA. The upper limit on the sum of the three previous eROSITA surveys is 0.6$\times 10^{-14}$~erg/s/cm$^2$ (0.3--2.2~keV). The upper limit on X-ray flux in the 3-sky review is 1.4$\times 10^{-14}$~erg/s/cm$^2$. Thus, the source became more than 8 times brighter in X-rays between the third and fourth surveys. This significant semiannual variability allows us to assign the source to a sample of eROSITA X-ray transients and to conduct more detailed multi-wavelength studies of the source for its classification. 

\section{Observations and data reduction}
\label{section_obs}

\subsection{Photometry} 

The first optical images of e2149 were acquired on 2021/10/10 in the $i$-SDSS filter on the 6-meter telescope BTA of the SAO RAS  using the SCORPIO-2 multimode focal reducer \citep{Afan}. The E2V~CCD261-84 detector in $2\times2$ binning mode provided a field of view of $6.8'$ with a sampling of 0.4$''$ per pixel. Six exposures of 15 seconds were obtained. An optical companion whose brightness varied from 23 to 21 mag was detected within the radius of R98. To understand the nature of the source, it was decided to perform its spectroscopy and long-term photometric observations. 

The photometric observations of e2149 were carried out on the 1.5m Russian-Turkish telescope RTT-150 (Turkish National Observatory T\"UB\'ITAK) from October 14 to November 16, 2021 and October 2-4, 2022. The TFOSC instrument and the $i$ filter of the SDSS system were used. Andor iKon-L 936 BEX2-DD-9ZQ CCD camera ($2048 \times 2048$ pixels) was used as a detector. The seeng was $1.5-2''$ at different observation periods, so a $2\times 2$ binning observation option and an image element of 0.65~$''$/pixel was used.  Additional photometric observations were carried out on November 5, 2021 on the 2.5m telescope of the Caucasus Mountain Observatory (CMO) of MSU in the $r$ band of the SDSS system in parallel with the spectral observations on the 6m telescope BTA. 
Measurements were made with a liquid nitrogen-cooled camera NBI\footnote{More information about the detector is available at: https://obs.sai.msu.ru/cmo/sai25/wfi/ .} (chip size is $4096 \times 4096$ pixels and image scale 0.155~$''$/pixel) mounted in the Cassegren focus of the telescope. The object was monitored for about 3 hours, with exposures of 300~sec. During the observations the seeng was 0.8--$1$~$''$ and no variations in atmospheric opacity were observed. The log of the photometric observations of e2149 is presented in Table \ref{log_phot}.

\begin{figure*}
  \centering
\includegraphics[width=12cm]{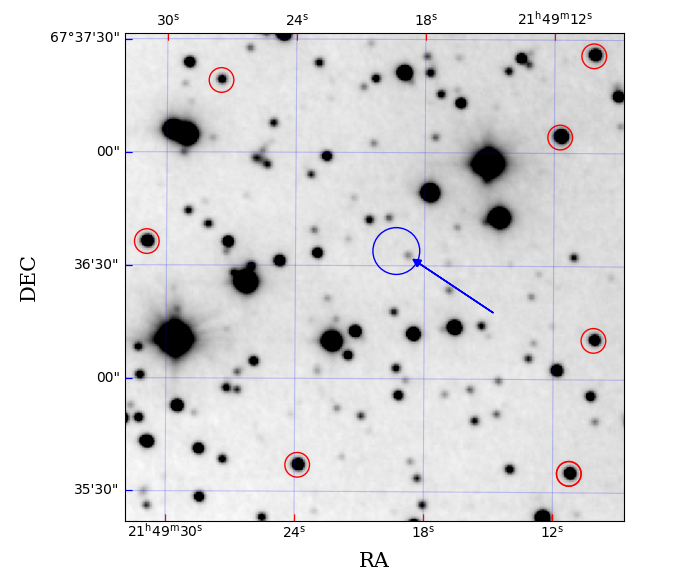}
\caption{The map of e2149 vicinity in the $i$ filter obtained by averaging the RTT-150 images. The blue circle with a radius of $5.5''$ shows the area of 98\% localization of the X-ray source. The center of the circle corresponds to the position of the X-ray source RA~$=327.33046^{\circ}$, DEC~$=+67.60941^{\circ}$. The arrow indicates an optical companion with coordinates RA~$=327.32866^{\circ}$, DEC~$=+67.60914^{\circ}$ for the epoch J2000. The red circles indicate the comparison stars used for differential photometry.}
\label{fig:findingchart}
\end{figure*}

The reduction of the obtained photometric data was carried out using the IRAF\footnote {The package for reduction and analysis of astronomical data was developed by the US National Optical Observatory and is available at: https://iraf-community.github.io/} package. Bias frames frames were subtracted from the images, multiplicative errors were reduced by the flat-field frames, and image cleaning from cosmic rays traces was performed using LaCosmic algorithm \citep{Dokkum}. The search for point-like sources in the images was performed using the DAOFIND algorithm. Aperture photometry of e2149 was carried out using several comparison stars shown in Fig. \ref{fig:findingchart}. Analysis of their relative brightness did not reveal their variability over the entire observation period. The optimal aperture size was found by minimizing the standard deviation of the brightness of the control stars comparable to e2149. 

\begin{table*}
\caption{The log of the photometric observations of e2149. The log lists telescopes used in observations, observation nights, observation period in heliocentric Julian dates, number of images acquired ($N$), filters used (C --- observations without filter), duration of single exposures ($\Delta t_{exp}$), total duration of observations in hours ($\Delta T$), and mean magnitude $\langle m \rangle$.}
\label{log_phot}
\begin{center}
\begin{tabular}{lccccccc}
\hline
Teleccope & Date           & Observation period,    & $N$	  & Filter & $\Delta t_{exp}$, \ & $\Delta T$, \ & $\langle m \rangle$,	\\ 
         &                & HJD-2459000          &     &        & sec & hours & mag \\ \hline
RТТ-150     & 14/15 Oct. 2021  & 502.32845--502.43391 & 19  & $i$     & 300 & 2.5 & 21.6 \\
RТТ-150     & 18/19 Oct. 2021  & 506.36633--506.47173 & 28  & $i$     & 300  & 2.4 & 21.8  \\
RТТ-150     & 19/20 Oct. 2021  & 507.21642--507.53612 & 60  & $i$     & 300 & 8 & 21.9 \\
RТТ-150     & 20/21 Oct. 2021  & 508.21643--508.63550 & 111 & $i$     & 300 & 10.6 & 22.1 \\
RТТ-150     & 28/29 Oct. 2021  & 516.23887--516.29336 & 14  & $i$     & 180, 300 & 1.7 & 22.8 \\
RТТ-150     & 05/06 Nov. 2021  & 524.20206--524.22764 & 4   & $i$     & 600 & 0.7 & 23.7  \\
RТТ-150     & 05/06 Nov. 2021  & 524.23487--524.36515 & 19  & C      & 600  & 3.1 & 23.7 \\
2.5-м CMO       & 05/06 Nov. 2021  & 524.23724--524.34754 & 22  & $r$      & 300  & 3 & 23.5  \\
RТТ-150     & 16/17 Nov. 2021  & 535.22147--535.26842 & 7   & C      & 600 & 1.1 & 23.2  \\
RТТ-150     & 02/03 Oct. 2022  & 855.31460--855.35774 & 12   & $i$       & 300 & 1.0 & 21.8  \\
RТТ-150     & 03/04 Oct. 2022  & 856.20068--856.33360 & 36   & $i$       & 300 & 3.1 & 22.0  \\
RТТ-150     & 04/05 Oct. 2022  & 857.32003--857.37078 & 11   & $i$       & 300 & 1.2 & 21.9  \\
\hline
\end{tabular}
\end{center}
\end{table*}

\subsection{Spectroscopy} 

The spectroscopic observations of e2149 were carried out on the 6m telescope BTA of the SAO RAS using the SCORPIO-2 multimode focal reducer in the long-slit spectroscopy mode \citep{Afan}.  
Observations were carried out on the night of November 5-6, 2021 under good atmospheric conditions (seeing $\approx 1''$). The VPHG1200@540 grism and $1.5''$-wide slit were used providing spectra in the range $\Delta\lambda = 3650-7250$~\AA~ with the  mean spectral resolution of $\delta \lambda \approx 6.5$~\AA. The slit had width of $6.8'$ providing the scale of $0.4''/$pixel.
Unfortunately, at the time of the observations of e2149 its brightness has weakened to $i \approx 23.5$~mag. Five spectra with a total exposure of 6000~sec were obtained.

The spectroscopic data were processed using a IDL package developed at SAO RAS for the reduction of long-slit spectra obtained by SCORPIO-2. The main processing steps are described in a number of our articles, e.g., in \citet{Egorov}. In contrast to the standard technique, the cleaning  of cosmic rays was carried out on individual frames already after the spectra had been calibrated to the wavelength scale and the spectrum of the night sky had been subtracted. This modifications are caused by relatively large size of the cosmic ray traces on the frames obtained by E2V~CCD261-84  detector \citep{AfanCCD}.  

Flux calibration was performed using the spectrum of the standard star BD+28$^{\circ}$4655, observed the same night before the observations of e2149. The extraction of the spectrum of e2149 was performed in a rectangular aperture of $5''$ width.

\section{Analysis of photometry}
\label{section_phot}

\begin{figure*}
  \centering
	\includegraphics[width=11cm]{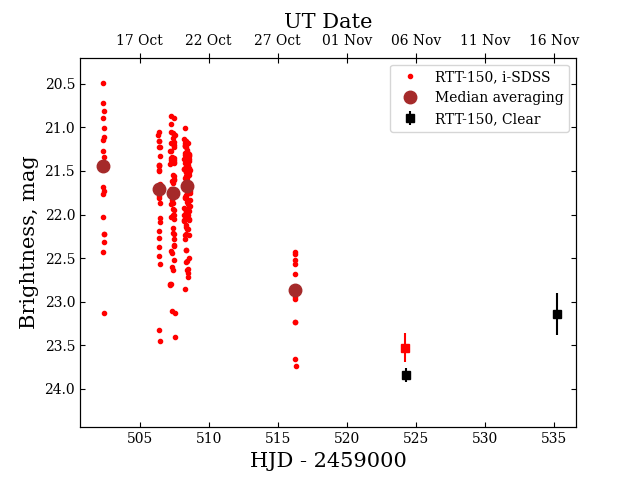}
\caption{Long-term light curve of e2149 based on observations with the RTT-150 telescope. Red dots are the light curves obtained in the $i$ filter between October 14 and 28, 2011, brown circles are  their median brightnesses. Black squares are estimates of the mean brightness of the star obtained without photometric filters on November 5 and 16. In addition, an estimate of the average brightness of e2149 in the $i$ band, found from the observations of November 5, is given.}
\label{fig:long}
\end{figure*}

The brightness of e2149 showed large changes during the photometric observations. Long-term light curve of e2149 obtained by observations at RTT-150 in 2021 is shown in Fig. \ref{fig:long}. It can be seen in the time period October 14-20, 2021, the average brightness of the object was at $\langle i \rangle =21.4-21.8$~mag and then dropped to the value $\langle i \rangle \approx 23.5$~mag recorded on November 5, 2021. e2149 was also observed in an intermediate state on October 28 with an average brightness $\langle i \rangle \approx 22.8$~mag. 
Table 1 shows that in the time period 2-4 October 2022, the source was again in the high state with an average brightness $\langle i \rangle \approx 21.9$~mag. 
In addition to the long-term variability the object shows short-period changes in brightness on time scales $\sim 1$~hour. 

\begin{figure}
  \centering
	\includegraphics[width=\columnwidth]{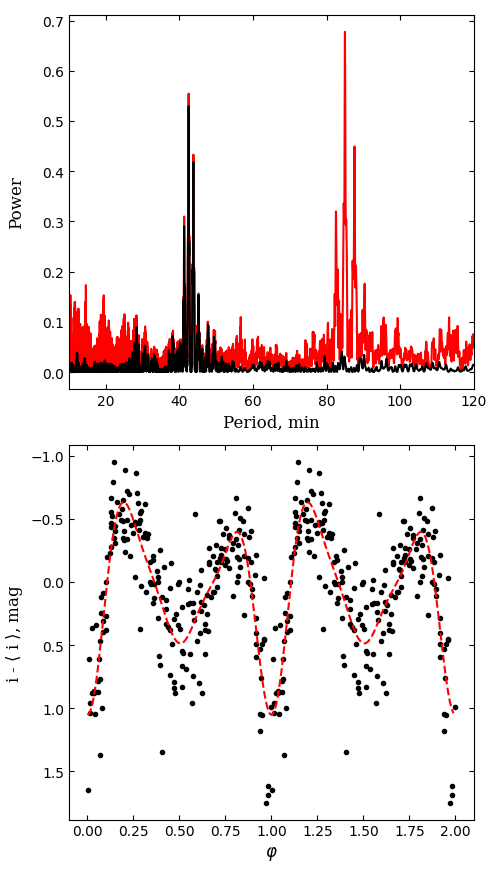}
\caption{The top panel shows the Lomb--Scargle periodograms of e2149 in high and intermediate states (14--28 October 2021). The black curve is the power spectrum obtained by one harmonic, the red curve is the power spectrum obtained by four harmonics. The lower panel shows the light curve of the high state of e2149 plotted in the phases of the photometric period according to the ephemeris (\ref{ephem}). The dotted line is a fit by trigonometric polynomial with four harmonics.}
\label{fig:ls}
\end{figure}

The analysis of short-term brightness variations of e2149 was carried out by Lomb-Scargle technique \citep{VanderPlas18}. The Lomb-Scargle periodograms were calculated using observations obtained during the time period October 14--28, 2021 when the object was in the high and intermediate states. The average light level was subtracted from the light curves of individual nights and the observations with poor detection of e2149 were excluded from consideration. Variability analysis based on a single harmonic ($\cos (\omega t)$) yielded a significant power peak corresponding on the period $P \approx 42.5$~min (see Fig. \ref{fig:ls}). However, a more reliable period occures when two or more harmonics are used. Fig. \ref{fig:ls} shows a Lomb-Scargle periodogram derived by four harmonics ($\cos (i \omega t)$, $i=1, \ldots, 4$) that provide a satisfactory (within errors of observations) fitting of the folded light curve. The power maximum corresponds to the period $P=85.0 \pm 0.4$~min.

There are also physical constraints which confirm the last period. The first period ($P \approx 42.5$~min) is significantly less than the minimal period for cataclysmic variables $P_{min} \approx 82$~min \citep{Knigge11}, and, if it corresponded to orbital variability, would indicate that e2149 belongs to AM~CVn type variables. However, this is in contradiction with the manifestation of the H$\alpha$ line in the spectra of e2149 (see below), which, like other hydrogen lines, is not observed in AM~CVn type stars. One could also assume that the observed periodicity is related to the accretor rotation, and the orbital period $P \gtrsim P_{min}$, i.e., e2149 is an intermediate polar. However, this would make it difficult to interpret the high brightness amplitude $\Delta i \approx 1.5$~mag, which is not typical for systems of such type. 

The found period is close to the $P_{min}$ and, on the other hand, close to the most probable orbital period for a cataclysmic variable (i.e., corresponding to the maximum of the observed distribution of cataclysmic variables \citep{Gansicke09}). The light curve of e2149 folded with the found period is shown in Fig. \ref{fig:ls}. It has a complex shape with two minima that differ in depth. The ephemerides of the main minimum in the light curve of e2149 are
\begin{equation}
    BJD_{min} = 2459502.3339(6) + 0.0590(2) \times E.
\label{ephem}
\end{equation}

\begin{figure*}
  \centering
	\includegraphics[width=\textwidth]{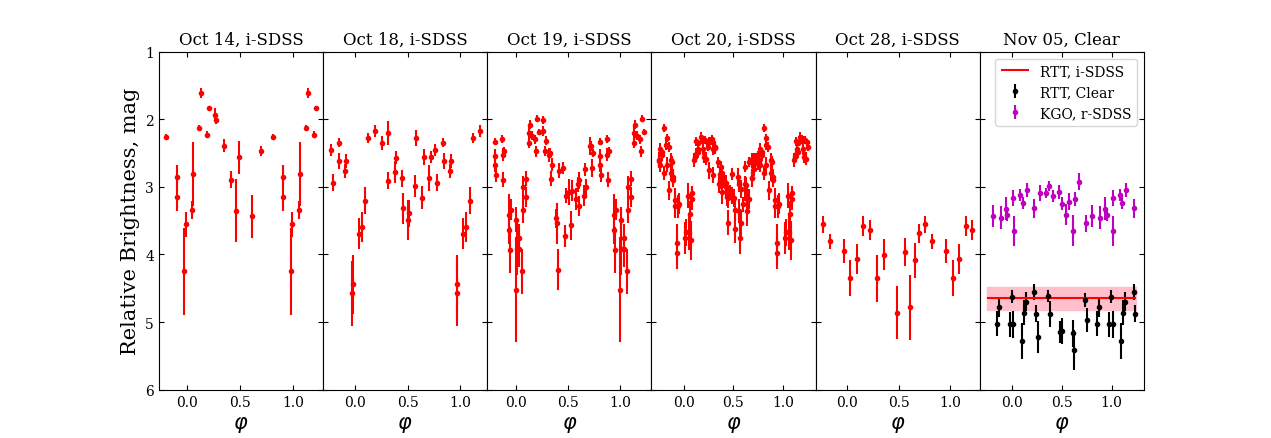}
\caption{Light curves of e2149 in the filter $i$ (red dots) obtained in the period between October 14 -- October 28, 2021. The last panel gives the light curve obtained on November 5 without the use of photometric filters (black dots). The red horizontal line represents the average brightness of the object in the $i$ band. There is also shown the light curve of e2149 obtained on the 2.5-m telescope CMO in the $r$ filter of the SDSS system. This light curve was not calibrated by absolute fluxes and is arbitrarily shifted along the ordinate axis.}
\label{fig:lcvar}
\end{figure*}

Let us consider in detail the folded light curve in Fig. \ref{fig:ls}. The center of the main minimum located at phase $\varphi=0$ is narrower ($\Delta \varphi \approx 0.2$) and has a sharp ingress and egress. The secondary minimum located at phase $\varphi\approx0.5$ is less deep and has a smooth ingress and egress. The change in the light curve of e2149 during the observations is shown in Fig. \ref{fig:lcvar}. It is evident that the brightness amplitude decreases with a decrease in the average brightness of the object. The highest amplitude was observed on October 14-20 ($\Delta i \approx 1.5-2$~mag) with an average brightness $i \approx 21.7$~mag (see Fig. \ref{fig:long}). Then, on October 28, the star's brightness decreased by $\approx 1$~mag and the amplitude dropped to $\Delta i \approx 1$~mag. In the low  state on November 5 observations of the variability of e2149 on RTT-150 were possible only without photometric filters. The brightness changes are not detectable on the obtained light curve and we can only impose restrictions on the amplitude $\Delta m \lesssim 0.5$~mag. On the same night e2149 was observed on the 2.5m telescope of the Caucasus Mountain Observatory of Moscow State University in the $r$ filter of the SDSS system. In contrast to the e2149 behavior in the high state in the $i$ band, the obtained light curve is a single-humped with an amplitude of $\Delta r \approx 0.5$~mag. Note that due to the error in the found period, the light curves from November 5 may be shifted in phase by $\Delta \varphi \approx 0.5$.

The light curve of e2149 in the high state resembles behavior of some polars. Examples are the polars BS~Tri \citep{Kolbin22}, EP Dra \citep{Schwope97}, V379~Vir \citep{Debes06}.  Their brightness minimum is due to the accretion spot position behind the observed stellar disk, and a double-humped maximum is formed during the spot's passage across the stellar disk. The double-humped light curve is due to the angle depended features of cyclotron radiation. Under certain physical conditions of the emitting medium, its intensity is maximal at angles between the magnetic field and line of sight $\theta \sim 90^{\circ}$ and decreases with decreasing $\theta$. Probably a similar scenario is realized in e2149, but it differs from the presented examples by a narrower minimum. The latter can be associated with a smaller inclination of the white dwarf rotation axis to the line of sight and, as a consequence, a shorter duration of the accretion spot behind the observed stellar disk (assuming that the spot is near the pole of rotation facing the observer). In the low state the accretion rate in the system is probably weakened. This leads to a decreased brightness of the accretion spot with a corresponding decrease in the system's brightness amplitude.

Unfortunately, the low brightness of e2149 during its spectral observations ($i \approx 23.5$~mag) did not allow us to obtain spectra suitable for estimating the system parameters (stellar components masses, orbital inclination) and the magnetic field strength of the white dwarf. Knowledge of these parameters would have allowed us to model the light curves with the determination of the magnetic dipole orientation and the position of the accretion spot. In this work, we limit ourselves to interpreting the light curve of e2149 in the high state within the model of an accreting magnetic white dwarf with dominated radiation from the accretion spot in the optical range. To perform this work, we used the code for calculating and approximating polar light curves described in \cite{Kolbin20}. This code uses a simple model of a white dwarf with a dipole magnetic field. The accretion spot is assumed to be geometrically-thin and is drawn on the surface of the star by magnetic field lines crossing the ballistic trajectory of the accretion stream. The accretion spot is assumed to be homogeneous in temperature and density. Its radiation intensity is calculated under the assumption of high Faraday rotation, which reduces the calculation of the Stokes parameters to two independent equations for the ordinary and extraordinary waves. The main contribution to the accretion spot radiation in the optical range belongs to the cyclotron radiation. The absorption coefficients calculated using the method of \cite{Chan81} were used to calculate radiation intensity. The accretion spot was assumed to be a point plane-parallel source, i.e., the depth of the spot along the line of sight $\ell$ behaves as $\ell\sim 1/\cos \gamma$, where $\gamma$ is the angle between the normal to the star surface and the line of sight.

\begin{figure}
  \centering
	\includegraphics[width=\columnwidth]{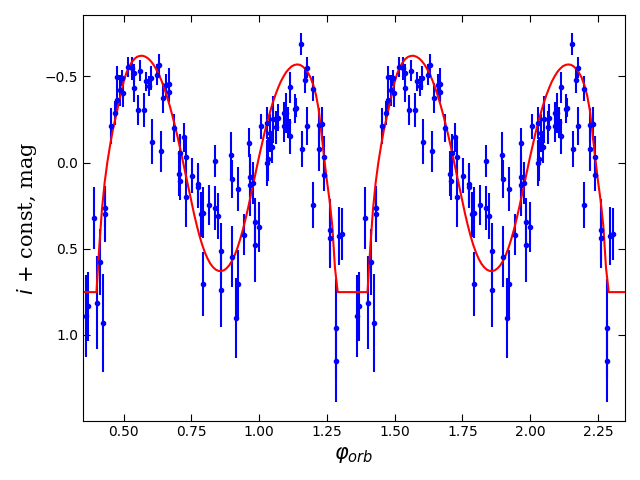}
\caption{The comparison of the observed (points with photometric error bars) light curve of e2149 with the theoretical one (continuous line). The light curves are given in the phase scale of the orbital period of e2149 (the phase $\varphi_{orb} = 0$ corresponds to the largest distance of the white dwarf from the observer).}
\label{fig:lcmod}
\end{figure}

\begin{figure}
  \centering
	\includegraphics[width=6cm]{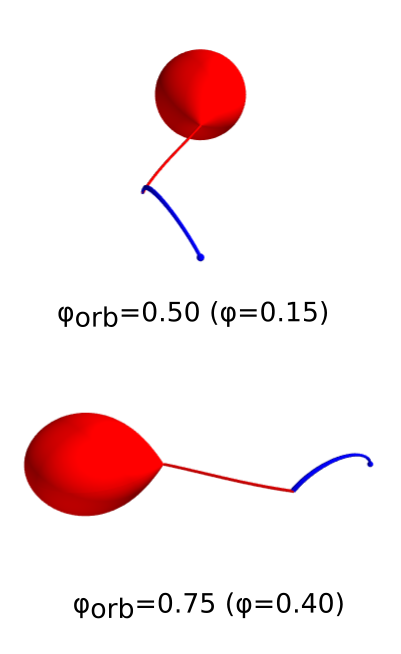}
\caption{The 3D-model of e2149 as seen from the observer in two phases of the orbital period. The red line shows the ballistic trajectory of the accretion stream flowing from the donor filling its Roche lobe. It changes to the magnetic trajectory (blue line), which ends at the surface of the white dwarf. The orbital period phases $\varphi_{orb}$ are indicated, as well as the photometric phases $\varphi$ at which $\varphi=0$ corresponds to the middle of the main minimum in the light curve.}
\label{fig:model}
\end{figure}

Using the described approach, we modeled the light curve of e2149 obtained on October 20, 2021. The code of \cite{Kolbin20} requires the masses of stellar components to calculate the ballistic trajectory. The mass of the white dwarf was assumed to be $0.83M_{\odot}$, i.e. the average mass of the accretor in cataclysmic variables \citep{Zolot11}. The mass of the secondary was taken equal to $0.07M_{\odot}$ \citep{Knigge11}. The inclination of the orbital plane was fixed at the value $i=60^{\circ}$. The cyclotron radiation intensity drops rapidly after the frequency $\approx 10 \omega_c$, where $\omega_c = eB/m_e c$ is the cyclotron frequency \citep{Wada80}. This makes it possible to impose a constraint on the magnetic field strength $B \gtrsim 10$~MG, since the cyclotron radiation variability is observed in the $i$ band. We used the magnetic field strength $B=20$~MG in modeling. The temperature of the accretion spot was assumed to be $20$~keV. To calculate the intensity of the spot radiation the plasma parameter $\Lambda = \omega_p^2 H / \omega_c c$ is also required, where $\omega_p$ is the plasma frequency and $H$ is the depth of the spot. This parameter was fixed at the value $\Lambda=10^3$ which is typical for polars. The orientation of the magnetic dipole and the position of the stagnation region (i.e., the region of transition from ballistic to magnetic trajectory), which determines the spot coordinates, were found by fitting the light curve using the least-squares technique. A genetic algorithm was used to minimize $\chi^2$ (see, e.g., \cite{Charb80}). As a result, the magnetic dipole orientation parameters were found to be $\beta = 22^{\circ}$, $\psi = 70^{\circ}$, where $\beta$ is the inclination of the dipole axis to the rotation axis, and $\psi$ is the magnetic pole longitude counted from the direction to the secondary in the orbital motion direction. The stagnation region is located at the azimuthal angle $\alpha=30^{\circ}$, counted from the direction to the secondary component. A comparison of the theoretical light curve with the observed one is given in Fig. \ref{fig:lcmod}. The observations are fitted with $\chi^2_{\nu}\approx 11$, and the deviation of $\chi^2_{\nu}$ from unity is due to the variable shape of the light curve during the observations, as well as possible rapid changes in brightness --- flickering. A three-dimensional model of the system visible in the two phases of the orbital period is shown in Fig. \ref{fig:model}. Although the presented model may be far from the real picture due to the large number of used assumptions, the present example shows that the shape of the light curve can be interpreted by the model of a magnetic white dwarf with a cyclotron radiation source and supports our assumption that the object belongs to AM~Her type systems.

\section{Spectra analysis}
\label{section_spec}

\begin{figure*}
  \centering
\includegraphics[width=13cm]{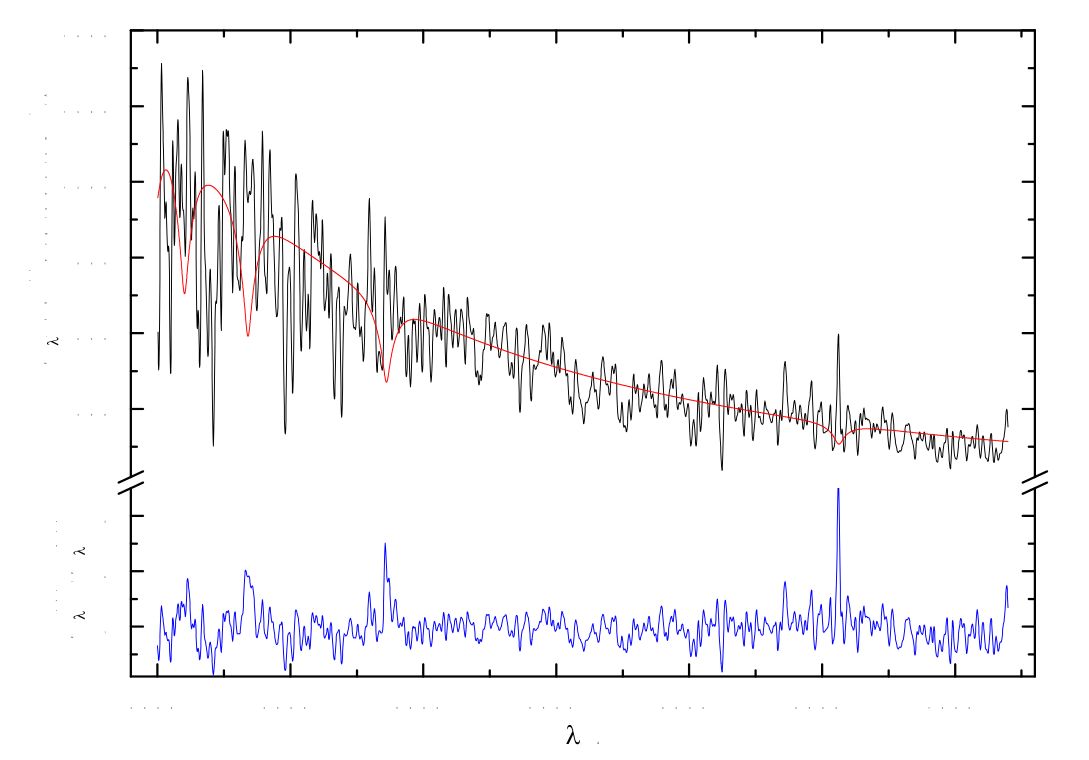}
\caption{The observed spectrum of e2149 (black line) and the model spectrum of a white dwarf with $T_{eff}= 24000$~K (red line). Blue line is the ratio of observed and theoretical fluxes.}
\label{fig:spec}
\end{figure*}

The averaged spectrum of e2149 is shown in Fig. \ref{fig:spec}. It has been smoothed with a Gaussian of FWHM~$=10$~\AA\, to reduce the noise. In addition, the fluxes were corrected for interstellar absorption usng $E(B-V) = 0.57$~mag in the direction of e2149. The resulting energy distribution has a monotonic slope to the red region ($dF/d\lambda \approx -2.6 \times 10^{-5}$erg/cm$^2$/\AA$^2$/sec). There is an emission peak at the wavelength $\lambda = 6561.1$~\AA\ with the intensity 3.7 times higher than the noise level in its vicinity. The half-width of the peak $\Delta\lambda \approx10$ \AA\, corresponds to the spectral resolution. This emission is probably the H$\alpha~\lambda$6562.8~\AA\ line shifted by a radial velocity of $ -80$ km/s.

The observed spectral energy distribution qualitatively corresponds to the thermal distribution at the temperature of the radiating medium above $T_{eff} = 20000$~K. We performed its fitting by theoretical spectra of white dwarfs with different values of $T_{eff}$. The $STAR$ program \citep{menzh14} and plane-parallel models of white dwarf atmospheres with radiative and convective energy transport obtained in \cite{mitr1} were used to calculate the model spectra. The abundances of all chemical elements except for hydrogen were assumed to be zero, and the line profiles of the Balmer series were modeled using the theory of \cite{vida1}. The spectra were calculated for a set of white dwarf atmospheres with $T_{eff}= 20000-40000$~K at a fixed value of gravitational acceleration $\log g = 8.3$ approximately corresponding to the white dwarf mass $M = 0.8 M_\odot$. The obtained spectra were convolved with the spectrograph instrumental function and compared with the observed one by calculating the standard deviation of the flux ratios
\begin{equation}
\sigma = \sqrt{\frac{1}{\lambda_2 -\lambda_1}\int \limits_{\lambda_1}^{\lambda_2} \left(\frac{F_{\lambda}^{obs}}{F_{\lambda}^{mod}} -1 \right)^2 d\lambda},
\end{equation}
where $F_{\lambda}^{obs}$ and $F_{\lambda}^{mod}$ are the fluxes in the observed and theoretical spectra. The range $\lambda\lambda = 4400-6700$~\AA\, with lower noise levels, was used in the comparison, excluding the H$\alpha$ and H$\beta$ line locations of $\Delta\lambda = 50$~\AA\ width. The pointed exceptions were made due to the possible emissions in hydrogen lines, uncertainties in the choice of $\log g$ values, and possible Zeeman splitting. The comparison of the model and observed spectra gives the temperature in the range $T_{eff}= 21000-30000$~K with the minimum of $\sigma$ at $T_{eff}= 24000$~K.

\section{Conclusions}

We performed optical study of the X-ray source e2149 detected by the eROSITA telescope in the 4th scan of the SRG all-sky X-ray survey. The photometric observations of e2149 reveal two brightness states differed by $\Delta i \approx 2$~mag. In the high state the source had double-peaked light curves in $i$-band with an amplitude of $\Delta i =1.5-2$~mag. In the low state the light curves have a smaller amplitude ($\Delta r \approx 0.5$~mag) while the spectra show a blue continuum with H$\alpha$ emission line ($EW = 18 \pm 5$~\AA). The found photometric period $P = 85.0 \pm 0.4$~min is close to the minimal orbital period $P_{min} \approx 82$~min for cataclysmic variables \citep{Knigge11}. On the other hand, periods close to $P_{min}$ are not rare among cataclysmic variables and, moreover, correspond to the maximum of their distribution \citep{Gansicke09}.

Based on the above observational features we can assume that e2149 belongs to magnetic cataclysmic variables of the AM~Her type (or polar). Rapid brightness state switches of several magnitudes in the long-term light curves of such systems are associated with a changes in matter transfer rates through the Lagrangian point L$_1$. Probably it is related to magnetic activity of the donor \citep{King98}. We have shown that the two-humped shape of the e2149 light curves in the high state can be described by the model of a magnetic white dwarf with a source of cyclotron  radiation (an accretion spot). The high brightness amplitude of $1.5-2$~mag is also typical for polars with a dominant contribution of the accretion spot in optical radiation. In the low state the spectrum of e2149 is consistent with a white dwarf of temperature $T_{eff}=24000$~K which would be expected for a reduced accretion rate. The unambiguous classification of e2149 as a polar can be made by polarimetric observations or spectral observations with the detection of cyclotron line harmonics.

{\bf Acknowledgements} The authors are grateful to TUBITAK, IKI, KFU, and the Academy of Sciences of the RT for partial support in the use of the RTT-150 (Russian-Turkish
1.5-m telescope in Antalya). Observations on the telescopes of SAO RAS are supported by the Ministry of Science and Higher Education of the Russian Federation. The upgrade of the telescope base is carried out within the framework of the national project "Science and Universities". \\
Part of the study was supported by the Russian Foundation for Basic Research under the scientific project \textnumero~19-32-60048. 
The photometric observations on the 1.5-m telescope RTT-150 and the 2.5-m telescope of the CMO GAISH MSU, and their initial processing were supported by grant RFF 21-12-00210 (IFB, ENI, EAS, SGJ).
The work of NAS, RIG, and IMH was funded by grant N 671-2020-0052 from the RF Ministry of Education and Science, allocated to Kazan Federal University to fulfill the state assignment
in the field of scientific activity. \\
This study is based on observations of the eROSITA telescope aboard the SRG observatory. The SRG observatory was manufactured by Roscosmos on behalf of the Russian Academy of Sciences, represented by the Institute for Space Research (IKI), as part of the Russian Federal Science Program with the participation of the German Aerospace Center (DLR). The SRG/eROSITA X-ray telescope was manufactured by a consortium of German institutes led by the Max Planck Society Institute for Extraterrestrial Astrophysics (MPE) with the support of DLR. The SRG spacecraft was designed, manufactured, launched, and operated by the Lavochkin NPO and its subcontractors. Science data are received by a complex of long-range space communications antennas in Bear Lakes, Ussuriysk, and Baikonur and are funded by Roscosmos. The eROSITA telescope data used in this work are processed using the eSASS software developed by the German eROSITA consortium and the software developed by the Russian SRG/eROSITA telescope consortium.

\end{document}